\documentstyle[epsf]{mn2e}

\newcommand{\kms}[0]{km\,s$^{-1}$}

\newcommand{\pc}[1]{\protect\citename{#1}}

\newcommand{\mlu}{M$_\odot$/L$_{B,\odot}$}
\newcommand{\ml}{M$_*$/L$_B$}

\begin{document}

\title[Redshift of the ring in MG1549+305]{Redshift of the Einstein
Ring in MG1549+305}

\author[Treu \& Koopmans]{T. Treu$^1$ \& L.~V.~E.~Koopmans$^{2,3}$ \\ 
$^1$ California Institute of Technology, Astronomy, mailcode 105-24,
Pasadena CA 91125, USA \\ 
$^2$ California Institute of Technology, Theoretical Astrophysics,
mailcode 130-33, Pasadena CA 91125, USA\\
$^3$ Present address: Space Telescope Science Institute, 3700 San Martin Drive, 
Baltimore MD 21218}

\pubyear{2003}

\label{firstpage}

\maketitle

\begin{abstract}
A deep spectrum taken with the Echelle Spectrograph and Imager (ESI)
at the Keck~II Telescope as part of the Lenses Structure and Dynamics
(LSD) Survey reveals the redshifts of the extremely red source of the
radio Einstein Ring in the gravitational lens system MG1549+305
($z_{\rm s}=1.170\pm 0.001$) and an intermediate redshift lensed
spiral galaxy ($z_{\rm G2}=0.604\pm 0.001$).  The source redshift
allows us to determine the mass of the SB0 lens galaxy enclosed by the
Einstein Radius ($R_{\rm E}=1\farcs15\pm0\farcs05$) $M_{\rm
E}$$\equiv$$M(<R_{\rm E})\,=\,8.4\pm0.7\times
10^{10}\,h_{65}^{-1}$~M$_\odot$. This corresponds to a Singular
Isothermal Ellipsoid (SIE) velocity dispersion $\sigma_{\rm
SIE}=214\pm5$~\kms, in good agreement with the measured stellar
velocity dispersion $\sigma=227\pm18$ \kms\, (Leh\'ar et al.\ 1996).
The mass-to-light ratio within the Einstein Radius ($\sim$1.4
effective radii) is $10\pm1\,h_{65}$ \mlu. This is only marginally
larger than typical stellar mass-to-light ratios of local early-type
galaxies, indicating that dark matter is not likely to be dominant
inside the Einstein Radius.
\end{abstract}

\begin{keywords}
gravitational lensing --- distance scale ---
galaxies: kinematics and dynamics --- 
galaxies: fundamental parameters ---
galaxies: elliptical and lenticular, cD
\end{keywords}

\vspace{-0.3cm}
\section{Introduction}

The number of known gravitational lens systems has grown steadily
since their initial discovery (Walsh et al. 1979), amounting now to
nearly a hundred
systems\footnote{e.g. http://cfa-www.harvard.edu/glensdata/
(CASTLES)}.

In addition to the pure theoretical interests in certain properties of
gravitational lensing (e.g. scaling relations, image multiplicities,
catastrophe theory, etc.) gravitational lenses have proven extremely
valuable for a variety of applications: (i) the determination of
cosmological parameters via lens statistics of well-defined samples
(e.g. Turner, Ostriker \& Gott 1984; Fukugita et al.\ 1992; Kochanek
1996; Falco, Kochanek \& Munoz 1998; Helbig et al. 2000; Sarbu, Rusin
\& Ma 2001; Chae et al.\ 2002), (ii) the study of the evolution of the
stellar populations of early-type galaxies (Keeton, Kochanek \& Falco
1998; Kochanek et al.\ 2000; Rusin et al.\ 2003; Treu \& Koopmans
2002; Koopmans \& Treu 2003; van de Ven, van Dokkum \& Franx 2003),
(iii) the determination of the Hubble constant (H$_0$) from
gravitational time-delays (e.g. Kundic et al. 1997; Schechter et al.\
1997; Koopmans \& Fassnacht 1999; Fassnacht et al.\ 1999; Koopmans et
al.\ 2000; Burud et al.\ 2002a, 2002b; Hjorth et al.\ 2002).

Unfortunately, the sample of known lenses is still plagued by
incomplete information on the redshift of the lens ($z_l$) or of the
source ($z_s$), which set the physical scale\footnote{The by-now well
established cosmological parameters (e.g. Spergel et al.\ 2003) affect
the physical scale at a negligible level for a given value of the
Hubble Constant.} of the system and is therefore vital for most
astrophysical applications of gravitational-lens systems.  This has
motivated dedicated efforts on large telescopes to complete the
determination of redshifts of the sample of known lenses
(e.g. Fassnacht \& Cohen 1998; Lubin et al. 2000; Tonry \& Kochanek
2000; Cohen, Blandford \& Lawrence 2002; McKean et al. 2003,
submitted).

Here, we present the redshift of the Einstein Ring in MG1549+305. The
redshift $z=1.170\pm 0.001$ of the extremely red object (ERO) source
of the radio ring was obtained at the Keck--II Telescope as part of
the Lenses Structure and Dynamics survey (Koopmans \& Treu 2002, 2003;
Treu \& Koopmans 2002a; hereafter KT02, KT03, TK02a). The currently
ongoing LSD survey aims at studying the mass distribution of
intermediate redshift ($0.1<z<1$) E/S0 galaxies by combining lensing
and dynamical analysis. The joint analysis -- based on HST images and
deep Keck spectroscopy -- allows us to remove degeneracies inherent to
each method alone and therefore to probe the mass distribution with
greater accuracy than is possible with either method individually
(e.g. Treu \& Koopmans 2002b).

The gravitational lens system MG1549+305 is potentially very useful
for studying the structure of early-type galaxies: the radio ring
(Leh\'ar et al. 1993) -- one of the handful of rings known -- provides a
wealth of information on the inner mass distribution of the
lens. Furthermore, the lens is relatively nearby ($z_l=0.111$; Leh\'ar
et al. 1993) and extended and is therefore ideally suited to study in
great detail the structure of the lens galaxy.  Unfortunately, the
redshift of the extremely red source has proven hard to measure,
preventing the use of this system in detailed analyses of early-type
lens galaxies (e.g. van de Ven et al.\ 2003 did not include it in
their sample). To allow for a prompt scientific exploitation of this
lens system we publish the redshift before our full lensing and
dynamical analyses of this system.

Throughout the paper, we use ${\rm H}_0=65$~km\,s$^{-1}$\,Mpc$^{-1}$,
and we assume $\Omega_{\rm m}=0.3$ and $\Omega_{\Lambda}=0.7$.

\begin{figure}
\begin{center}
\leavevmode
\hbox{%
\epsfxsize=1.05\hsize
\epsffile{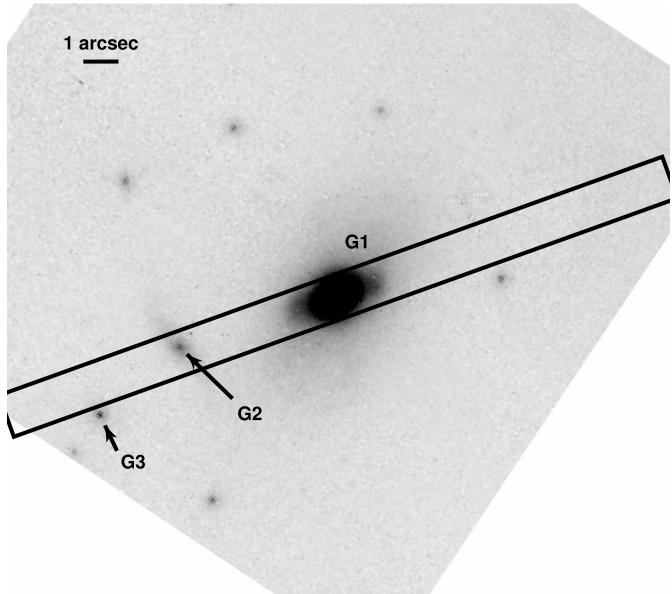}}
\end{center}
\caption{NICMOS image of MG1549+305 through filter F160W. North is up,
East is Left. The image is $\approx19''$ on a side. The galaxies are
labelled following Leh\'ar et al.\ (1993): G1 is the lens, G2 is an
intervening late-type galaxy and G3 is the source of the radio
ring. The spectroscopic slit used in the observations is
overplotted. At ground based resolution (seeing FWHM $0\farcs6$) a
non-negligible fraction of the light from G3 falls in the slit,
yielding the spectrum shown in Figure~3.}
\label{fig:images}
\end{figure}

\vspace{-0.3cm}
\section{Observations}

Spectroscopy of MG1549+305 (Figure~1) was performed on July 21--22
2001, using the Echelle Spectrograph and Imager (ESI; Sheinis et al.\
2002) at the Keck-II Telescope in Echelle mode. The slit
($1\farcs25\times20''$) was centered on the lens galaxy G1 (the
notation of Leh\'ar et al.\ 1993 is adopted throughout this paper), at
position angle PA\,$=110^\circ$, i.e. aligned with the bar. This
choice of PA yielded at the same time spectra of galaxies G2 -- an
intervening late-type galaxy -- and of the source of the radio ring,
galaxy G3. Five exposures totalling 7800s were obtained, dithering
along the slit by $5''$ after each exposure. The seeing was
$0\farcs6-0\farcs7$ and the nights were photometric. The spectra were
reduced using the ESI two-dimensional reduction {\sc iraf} package
EASI2D developed by D.~J.~Sand and T.~Treu (Sand, Treu, Ellis 2002;
Sand et al.\ 2003, in preparation). Relevant parts of the spectra of
galaxies G2 and G3 are shown in Figures~2 and 3.

\begin{figure}
\begin{center}
\leavevmode
\hbox{%
\epsfxsize=1.05\hsize
\epsffile{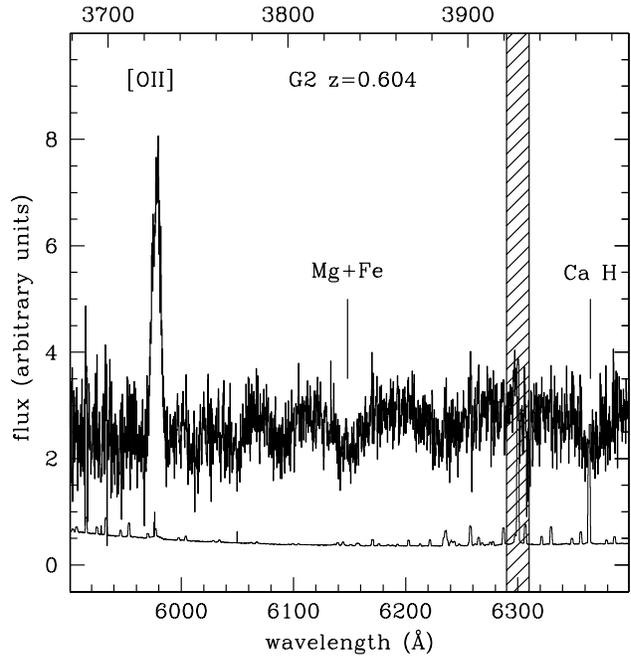}}
\end{center}
\caption{Part of the spectrum of Galaxy G2. The lower curve
represents the noise}
\label{fig:G2}
\end{figure}

\begin{figure}
\begin{center}
\leavevmode
\hbox{%
\epsfxsize=1.05\hsize
\epsffile{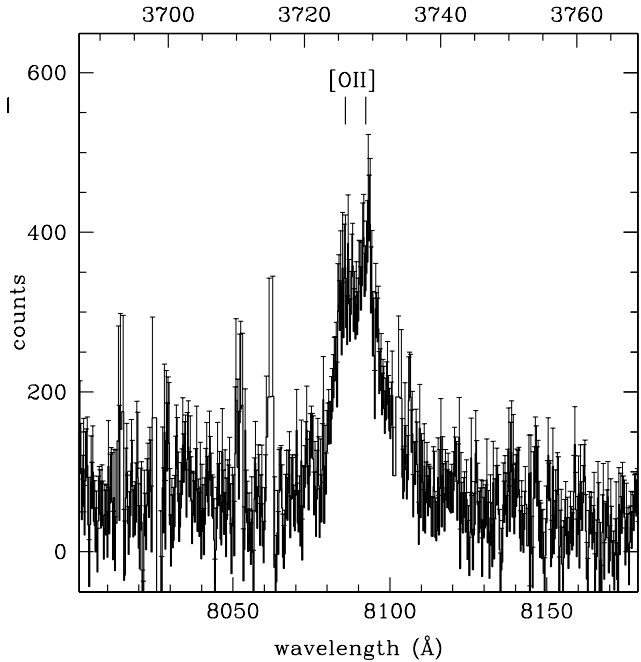}}
\end{center}
\caption{Part of the spectrum of the lensed source G3. The error bars
represent the noise. Note that, the spectral resolving power of ESI is
sufficient to resolve the doublet, with correct wavelength ratio,
which we identify as [OII]3726,3729.}
\label{fig:G3}
\end{figure}

Several features yield unambiguously the redshift of G2 ($z_{\rm
G2}=0.604\pm 0.001$; also [OIII] and H$\beta$ emission lines are
detected at the appropriate wavelengths). In contrast, only one
feature is detected for G3 in the entire spectral range 3,900--10,500
\AA. Since the line is a doublet with the correct separation we
identify it as [OII]3726,3729\AA\ at $z=1.170 \pm 0.001$.

To cross-check the redshift of G3 we measured the photometric
properties of galaxy G3. To this aim Hubble Space Telescope (HST)
images of MG1549+305 were taken from the HST archive and reduced as
described in KT02a. The Near Infrared Camera and Multiple Object
Spectrograph (NICMOS; Camera 2) image through filter F160W (hereafter
H$_{16}$; total exposure time 1111.58s; GO7495; PI: Falco) is shown in
Figure~1.  Photometry on the NICMOS image was measured using
SExtractor (Bertin \& Arnouts 1996), yielding a total magnitude ({\tt
mag\_auto}) of the source of the radio ring (galaxy G3) of
H$_{16}=19.6\pm0.1\pm0.1$ (random+systematic errors).  Unfortunately,
the Wide Field and Planetary Camera 2 (WFPC2) images through filters
F555W and F814W (total exposure time 640s on the planetary camera; GO7495;
PI: Falco) are rather shallow and galaxy G3 is undetected.

\begin{table}
\caption{Coordinates and redshifts of the galaxies in the MG1549+305
system. Offsets are relative to the position of G3, RA=15:49:12.87
DEC=30:47:12.90 (J2000). Uncertainties are $0\farcs02$ on offsets and
0.001 on redshifts.}
\label{tab:pos}
\begin{tabular}{lccc}
  \hline
Galaxy & $\Delta$RA(J2000) & $\Delta$DEC(J2000) & z \\
\hline
G1 &  $-6\farcs64$ & $3\farcs40$ & 0.111 \\ 
G2 &  $-2\farcs26$ & $1\farcs91$ & 0.604 \\ 
G3 &  $\equiv0$ & $\equiv0$ & 1.170\\
\hline
\end{tabular}
\end{table}

We obtained deeper V and I band images using ESI in imaging mode on
July 26 2001. The seeing was $0\farcs6$ FWHM, conditions were
photometric, and the exposure time totaled 300s per filter.
Photometric calibration, accurate to 0.05 mags, was obtained by
imaging photometric standards (Landolt 1992). The images were reduced
in a standard manner.  After matching the resolution of the NICMOS
image to the resolution of the ground based images, we measured the
colors of G3 to be V--H$=4.8\pm0.2$ and I--H$=3.1\pm0.1$ ($\pm0.1$
total systematic uncertainty on the zero point calibration of NICMOS
and ESI). The luminosity and colors of G3 are typical of radio
galaxies at $z\approx1$ (e.g. Best et al.\ 1998; de Breuck at al.\
2002) supporting the redshift identification.

\section{Discussion} 

\label{sec:disc}

With the redshift of the source ($z_{\rm s}=1.170 \pm0.001$), we can
derive some basic physical properties of the lens galaxy G1. Adopting
the Einstein Radius by Leh\'ar et al.\ (1993; $R_{\rm
E}=1\farcs15\pm0\farcs05$; see also Rusin et al.\ 2003), the projected
mass enclosed by the Einstein Radius is $M_{\rm E}=(8.4\pm 0.7)\times
10^{10}\,h_{65}^{-1}$~M$_{\odot}$ and a Singular Isothermal Ellipsoid
(SIE) velocity dispersion of $\sigma_{\rm SIE}=214\pm5$~\kms\ is
derived (Kormann et al.\ 1994).  Note that G2 is at an intermediate
redshift between the lens and the source and a detailed lens model
should therefore be done with at least two lens planes. However, G2 is
a relatively faint late-type galaxy $\sim$4\,$R_{\rm E}$ away from G1
and can, for modeling purposes, be regarded as a minor perturber at
the redshift of G1 (Kochanek \& Apostolakis 1988).

Using the effective radius ($0\farcs81\pm0\farcs08$), total magnitude
and colors tabulated by Rusin et al.\ (2003), we derive the magnitudes
of the lens galaxy within the Einstein Radius, F814W=$17.25\pm0.05$ and
F555W=$18.74\pm0.05$, assuming an $R^{1/4}$ surface brightness
profile. Correcting to rest frame (Treu et al.\ 2001) and for
foreground extinction (Schlegel et al.\ 1998), the luminosity within
the Einstein Radius is $L_B=(8.5\pm0.8)\times
10^9\,h_{65}^{-2}$~L$_{B,\odot}$.

The mass--to--light ratio enclosed by the Einstein Ring is then \ml$ =
(10\pm1)\,h_{65}$~\mlu, in the upper range of the typical stellar
values for local early-type galaxies [($7.3\pm2.2)\,h_{65}$~\mlu\ ;
TK02a and references therein], but not dramatically larger. This
suggests that dark matter is not dominant within the Einstein Radius,
as expected given that it is similar to the effective radius
(e.g. KT02b). To show the presence of dark-matter inside the Einstein
Radius more sophisticated dynamical and lens models are
required. Nevertheless, it is already remarkable that despite the
complex morphological inner structure of the barred lens galaxy G1,
the stellar velocity dispersion measured by Leh\'ar et al.\ (1996)
$\sigma=227\pm18$ \kms\, is in such good agreement with the equivalent
singular isothermal ellipsoid velocity dispersion $\sigma_{\rm
SIE}=214\pm5$\kms.

The system MG1549+305 -- with its detailed multi-colors HST images, a
radio Einstein Ring and extended kinematic profiles along major and
minor axes -- is an ideal case to explore how this ``coincidence''
arises. The structure of the radio Einstein ring provides unique
information on the projected mass distribution while its relatively
large angular size and luminosity makes it possible to derive accurate
and spatially extended kinematic information along more then one
axis. The combination of these constraints should allow us to
accurately measure its internal mass distribution and orbital
structure. A more detailed model of this system is being developed as
part of the LSD Survey.

\medskip

\section{Acknowledgments}
Based on observations collected with the NASA/ESA HST, obtained at
STScI, which is operated by AURA, under NASA contract NAS5-26555 and
at the W.M. Keck Observatory, which is operated as a scientific
partnership among the California Institute of Technology, the
University of California and the National Aeronautics and Space
Administration. The Observatory was made possible by the generous
financial support of the W.M. Keck Foundation. The ESI data were
reduced using software developed in collaboration with D.J.~Sand.  We
acknowledge the use of the HST data collected by the CASTLES
collaboration. LVEK and TT acknowledge support by grants from NSF and
NASA (AST--9900866; STScI--GO 06543.03--95A; STScI-AR-09960). We thank
J. Miller, M. Bolte, R. Guhathakurta, D. Zaritsky and all the people
who worked to make ESI such a nice instrument. Finally, the authors
wish to recognize and acknowledge the very significant cultural role
and reverence that the summit of Mauna Kea has always had within the
indigenous Hawaiian community.  We are most fortunate to have the
opportunity to conduct observations from this mountain. Finally, we
thank the anonymous referee for comments that improved the
presentation of the manuscript.

\end{document}